\documentclass[twocol]{ametsocV6.1}

\usepackage[utf8]{inputenc} % set input encoding (not needed with XeLaTeX)
\usepackage{url}
\usepackage[T1]{fontenc}

\usepackage{graphicx} % support the \includegraphics command and options
\usepackage{multirow}
\usepackage{amsfonts}
\usepackage{bbm}

\usepackage{booktabs} % for much better looking tables
\usepackage{array} % for better arrays (eg matrices) in maths
\usepackage{paralist} % very flexible & customisable lists (eg. enumerate/itemize, etc.)
\usepackage{verbatim} % adds environment for commenting out blocks of text & for better verbatim
\usepackage{subfig} % make it possible to include more than one captioned figure/table in a single float
\usepackage{amsmath}
\usepackage{authblk}
\usepackage{float}
\usepackage{textcomp}
\usepackage{siunitx}
\usepackage{tikz}
\usetikzlibrary{shapes.geometric, arrows, fit, topaths, positioning}
\tikzstyle{proc} = [rectangle, minimum width=2cm, minimum height=1cm, text centered, draw=black]
\tikzstyle{io} = [rectangle, rounded corners, minimum width=2cm, minimum height=1cm, text centered, draw=black, font=\bf]
\tikzstyle{block} = [rectangle, dashed, draw=black]
\tikzstyle{arrow} = [thick, ->, >=stealth]

\title{Statistical post-processing yields accurate probabilistic forecasts from Artificial Intelligence weather models}

\authors{Belinda Trotta\aff{a}, Robert Johnson\aff{a}, Catherine de Burgh-Day\aff{a}, Debra Hudson\aff{a}, Esteban Abellan\aff{a}, James Canvin\aff{a}, 
Andrew Kelly\aff{a}, Daniel Mentiplay\aff{a}, Benjamin Owen\aff{a}, Jennifer Whelan\aff{a}\correspondingauthor{Belinda Trotta, belinda.trotta@bom.gov.au}}

\affiliation{\aff{a}{Bureau of Meteorology, Australia}}

\abstract{Artificial Intelligence (AI) weather models are now reaching operational-grade performance for some variables, but like traditional Numerical Weather Prediction (NWP) models, they exhibit systematic biases and reliability issues. We test the application of the Bureau of Meteorology’s existing statistical post-processing system, IMPROVER, to ECMWF’s deterministic Artificial Intelligence Forecasting System (AIFS), and compare results against post-processed outputs from the ECMWF HRES and ENS models. Without any modification to processing workflows, post-processing yields comparable accuracy improvements for AIFS as for traditional NWP forecasts, in both expected value and probabilistic outputs. We show that blending AIFS with NWP models improves overall forecast skill, even when AIFS alone is not the most accurate component. These findings show that statistical post-processing methods developed for NWP are directly applicable to AI models, enabling national meteorological centres to incorporate AI forecasts into existing workflows in a low-risk, incremental fashion.}

\begin{document}

%% Necessary!
\maketitle

\section*{Notice}
This Work has been accepted by Artificial Intelligence for the Earth Systems. The AMS does not guarantee that the copy provided here is an accurate copy of the Version of Record (VoR).

\section{Introduction}\label{introduction}

Artificial Intelligence weather models have rapidly improved in the last few years \citep{BurghDay}. Deterministic AI models now typically achieve RMSE accuracy as good or better than traditional Numerical Weather Prediction (NWP) models for medium-range forecasting on most variables assessed, while being many times faster to run. In this study, we focus on the European Centre for Medium-Range Weather Forecasting's (ECMWF) deterministic Artificial Intelligence Forecasting System (AIFS) model \citep{aifs}, which 
is operationally supported as of 25 February 2025 \citep{aifsoperational}.  AIFS is a graph transformer neural network. Graph-based architectures have been hugely successful in weather forecasting. Deepmind's GraphCast model was found to have better RMSE skill score than ECMWF's HRES deterministic model in around 90\% of cases \citep{graphcast}. Like GraphCast, AIFS generally outperforms HRES \citep{aifs}.  Earlier work by \cite{kiesler} used a much smaller graph-based network and achieved results better than the Global Forecasting System (GFS, produced by the US National Oceanic and Atmospheric Administration) but not as good as the ECMWF's HRES model.

As is the case for traditional NWP models, the resolution of AI weather models is limited by computational constraints. The resolution of AIFS is approximately \ang{0.25} \citep{aifs}, significantly coarser than the \ang{0.1} resolution of ECMWF HRES \citep{hres}.  Furthermore, the graph-based models described above are all auto-regressive: the first timestep is calculated by running the model on the current analysis, and then subsequent timesteps take as input the model output of the previous timestep. Therefore,  model errors can accumulate over time. Thus, as with NWP models, a key objective for post-processing AI models is to downscale and bias-correct the forecast. Additionally, operational post-processing systems now commonly produce probabilistic as well as deterministic outputs, and combine various input models into a blended output forecast. We hypothesise that existing post-processing systems, developed for NWP models, can be used without modification to fulfill all these objectives for AI models.

Since AI weather models have only recently become good enough to be considered for operational use, the study of post-processing such models is relatively undeveloped, compared to the large body of work on NWP models.  However, existing research shows benefits from applying post-processing. \cite{bremnes} use a neural network to produce calibrated probabilistic site forecasts of wind speed and temperature for the Pangu-Weather deterministic ML forecast \citep{pangu}, and apply the same methods to some deterministic and ensemble NWP forecasts. They find that post-processing yields large improvements in accuracy for both AI and NWP forecasts.

\cite{bulte} evaluate two post-processing methods for producing probabilistic forecasts from Pangu-Weather, and compare the results with the ECMWF raw ensemble. In contrast to the work of \cite{bremnes}, the models are trained and evaluated on gridded analyses rather than site data.  The first method evaluated by \cite{bulte} is the EasyUQ technique developed by \cite{easyuq} and based on isotonic distributional regression of \cite{idr}. This is a non-parametric calibration method somewhat similar to the reliability calibration used in the present work (and described in more detail in Section \ref{methods}\ref{improver}). The second method is distributional regression networks, an approach where post-processed forecasts follow parametric distributions whose parameters are predicted by a neural network based on the inputs. While the latter method allows incorporating additional predictors, it is found not to offer significant advantages over the simpler method. Both methods allow the post-processed deterministic forecast to achieve similar or better accuracy than the raw ECMWF ensemble, at least for the first few days of the forecast period. A much more sophisticated method is demonstrated by \cite{ge}, who use a deep neural network to bias correct and downscale gridded predictions to the point scale. Since the network has billions of parameters, training is computationally intensive and requires a long history of training data.

Here we apply the IMPROVER \citep{improver} post-processing system to post-process the AIFS forecast outputs, and apply the same methods to ECMWF's deterministic HRES and ensemble (ENS) models to provide a comparison. We demonstrate that although IMPROVER is developed for post-processing traditional NWP forecasts, it is also effective for AIFS with no changes to the configuration parameters or processing workflows. We also show that adding AIFS to a blend of models with the two traditional NWP forecasts improves the accuracy of the blend, even in situations where AIFS is less accurate than one of the other models. One of the key features of IMPROVER is that it facilitates producing probabilistic forecasts from both deterministic and ensemble NWP inputs. The probabilistic forecasts produced by AIFS are of comparable quality to those produced by ECMWF HRES and including AIFS in the blend improves the quality of the probabilistic forecast.

\section{Data}\label{data}

The study uses data spanning 1 March 2024 to 23 July 2024 (with a few days of missing data, varying by parameter). The date range was limited by computational resources and the availability of archived pre-processed data. In future work, 
it would be interesting to consider a longer date range including the Southern hemisphere summer period. AIFS data was downloaded from ECMWF's Meteorological Archival and Retrieval System (MARS). Since calibration uses a rolling period of the previous 30 days of forecasts, the outputs are less accurate for the first month, and we analyse the results only from 1 April onwards, amounting to approximately 16 weeks of data. We consider forecasts from the 12Z basetime for each day in the period.\footnote{Basetime refers to the time that the model's initial conditions represent.} We selected this basetime since, of the Bureau's two daily operational forecasts, this one receives the most attention. We post-process 3 surface-level variables: temperature, dew point temperature, and wind speed at 10m. We evaluate both the expected value and probabilistic output forecasts.

Post-processed forecasts are produced for lead times up to 240 hours at 1-hour intervals. The raw AIFS forecast has lead times at 6-hourly intervals, while the ENS and HRES forecasts have 3-hourly frequency up to 150 hours, and 6-hourly thereafter. As described in Section \ref{methods}, the input forecasts are interpolated to the output frequency. The resolutions of the NWP forecasts ENS and HRES are \ang{0.2} and \ang{0.1}, respectively.\footnote{ENS is produced by ECMWF at \ang{0.1} resolution; the Bureau receives an upscaled version.} The AIFS forecast is produced on the N320 reduced Gaussian grid (see \cite{aifs}). Our bias correction and calibration use the gridded analysis of the Australian domain produced by the Mesoscale Surface Analysis System (MSAS) \citep{msas}, which has a resolution of 2.5 arc minutes (approximately 4 km for the region modelled).

\begin{figure*}
\begin{center}
\includegraphics[width=15cm]{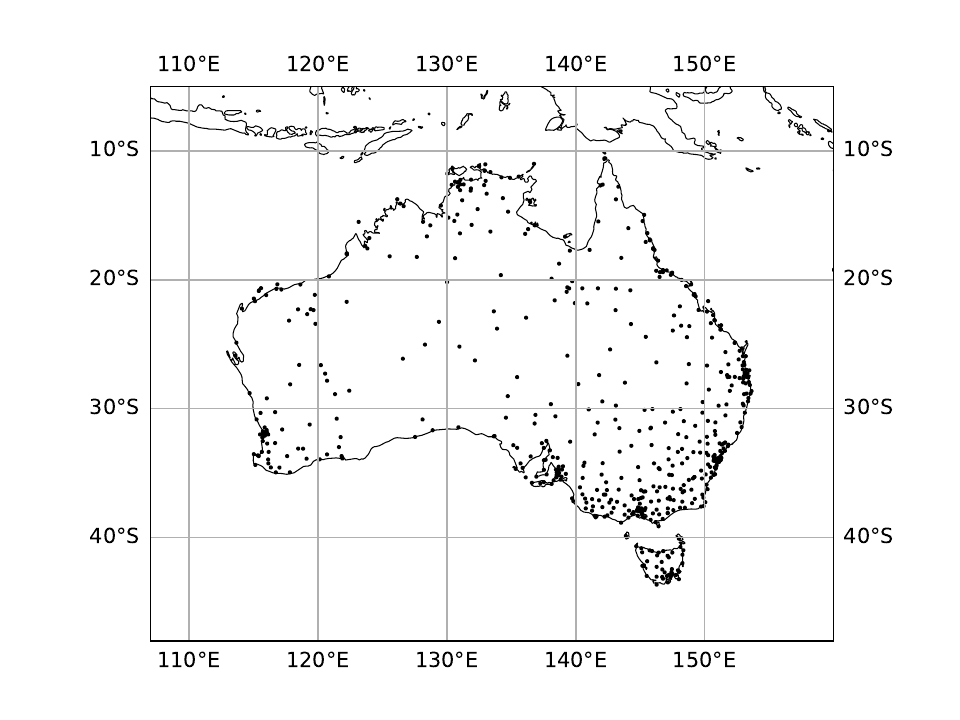}
\caption{The Australian continent, showing locations of the Bureau of Meteorology's Automatic Weather Stations used for verification in this study.}
\label{fig:map}
\end{center}
\end{figure*}

The ground truth for our verification is hourly observation data  extracted from the Bureau of Meteorology's Jive database \citep{jive}, which contains quality-controlled observations from the Bureau's network of Automatic Weather Stations. In total, 569 weather stations are used in the analysis. Figure \ref{fig:map} shows the location of these stations. Observation data is also used for optimising the blending weights as discussed in Section \ref{methods}\ref{blending}.

\section{Methods}\label{methods}

The full analysis workflow is summarised in Figure \ref{fig:flowchart}.

\begin{figure*}
\centering
\begin{tikzpicture}
\node[align=center] (input) [io, xshift=2cm] {Input};
\node[align=center] (ekr) [proc, below of=input, yshift=-0.5cm] {Regrid to lat/lon\\(AIFS only)};
\node[align=center] (imp) [proc, dashed, below of=ekr, yshift=-0.5cm] {IMPROVER processing};
\node[align=center] (imp_output)[io, below of=imp, yshift=-0.5cm] {Gridded\\probabilities};
\node[align=center] (extract) [proc, below of=imp_output, yshift=-0.5cm] {Extract sites};
\node[align=center] (siteprob) [io, below of=extract, yshift=-0.5cm] {Probabilities\\at sites};
\node[align=center] (calcexpval) [proc, below of=siteprob, yshift=-0.5cm, xshift=3cm] {Calculate\\expected values};
\node[align=center] (expval)[io, below of=calcexpval, yshift=-0.5cm] {Expected values\\at sites};
\node[align=center] (blendprob) [proc, below of=siteprob, yshift=-0.5cm, xshift=-3cm] {Blend probabilities\\with other forecasts};
\node[align=center] (blendexpval) [proc, below of=expval, yshift=-0.5cm] {Blend expected values\\with other forecasts};
\draw [arrow] (input) -- (ekr);
\draw [arrow] (ekr) -- (imp);
\draw [arrow] (imp) -- (imp_output);
\draw [arrow] (imp_output) -- (extract);
\draw[arrow] (extract) -- (siteprob);
\draw[arrow] (siteprob) -- (blendprob);
\draw[arrow] (siteprob) -- (calcexpval);
\draw[arrow] (calcexpval) -- (expval);
\draw[arrow] (expval) -- (blendexpval);
\node[align=center] (interp) [proc, right of=input, xshift=6.5cm, yshift=1cm] {Interpolate\\time steps};
\node[align=center] (regrid) [proc, below of=interp, yshift=-0.5cm] {Regrid};
\node[align=center] (lapse) [proc, below of=regrid, yshift=-0.5cm] {Lapse rate correction\\(temperature only)};
\node[align=center] (bias) [proc, below of=lapse, yshift=-0.5cm] {Bias correction};
\node[align=center] (prob) [proc, below of=bias, yshift=-0.75cm] {(Fuzzy) thresholding\\(transform forecast\\to probabilities)};
\node[align=center] (spatial) [proc, below of=prob, yshift=-0.75cm] {Spatial smoothing};
\node[align=center] (relcal) [proc, below of=spatial, yshift=-0.5cm] {Reliablity calibration};
\node[proc, fit=(interp)(regrid)(lapse)(bias)(prob)(spatial)(relcal), minimum height=11cm, minimum width=4.5cm] (impbox) {};
\draw[dashed] (imp.north east) -- (impbox.north west);
\draw[dashed] (imp.south east) -- (impbox.south west);
\draw[arrow] (interp) -- (regrid);
\draw[arrow] (regrid) -- (lapse);
\draw[arrow] (lapse) -- (bias);
\draw[arrow] (bias) -- (prob);
\draw[arrow] (prob) -- (spatial);
\draw[arrow] (spatial) -- (relcal);
\end{tikzpicture}
\caption{The processing workflow. The flowchart on the left shows the main steps of the analysis. These steps are duplicated for each input model (ENS, HRES, and AIFS). Inputs and outputs of selected steps are shown in bold. The IMPROVER processing is expanded in more detail on the right.}
\label{fig:flowchart}
\end{figure*}

AIFS data was pre-processed to a format expected by the IMPROVER software. As mentioned above, AIFS is produced on the N320 reduced Gaussian grid. The IMPROVER software cannot natively handle this grid, so as part of pre-processing we transform it to a latitude/longitude grid of resolution \ang{0.25}. IMPROVER is already in use operationally at the Bureau of Meteorology for HRES, ENS, and MSAS, so for these models pre-processed data was retrieved from an operational archive.\footnote{This pre-processing takes place prior to the IMPROVER workflow described below and is very minimal. It primarily consists of standardising file formats (including converting from GRIB to NetCDF format, standardising variable names, and extracting the Australian region).} The inputs and outputs for IMPROVER cover the Australian region only; for the global models HRES, ENS, and AIFS, the Australian domain is extracted in pre-processing that takes place before IMPROVER.

The IMPROVER software produces probabilistic forecasts by predicting the probabilities of exceedence at a set of thresholds. The thresholds are chosen so that they cover the range of forecast values one could reasonably expect for the Australian climate. We use quite a large number of thresholds so that the probability distributions can be modelled realistically: there are 61 thresholds for temperature, 47 for dewpoint, and 49 for windspeed. Expected value forecasts are calculated from the probabilistic output by linearly interpolating these probabilities and integrating the resulting distribution.

The Bureau's IMPROVER processing workflow is summarised in Section \ref{improver} and described in detail in \cite{improver-rr} and \cite{fuzzy}. Forecast bias and probability calibration are corrected using a rolling history of 30 days of data. The ground truth for 
these corrections is the gridded MSAS analysis mentioned above. Probability calibration is implemented using the reliability calibration algorithm described by \cite{flowerdew}.

After running the IMPROVER processing with AIFS, the output was extracted at the observation sites (Figure \ref{fig:map}), using the nearest-neighbour method.\footnote{The IMPROVER software was also used to extract the site forecasts, but this was achieved with a post-processing script separate from the main IMPROVER processing workflow. This is because the Bureau's operational IMPROVER workflow scheduler, which we adapted for this analysis, produces only gridded forecasts.} Expected values were calculated from the thresholded forecast.\footnote{Again, this was done using the IMPROVER software, but as a post-processing step separate from the main processing of gridded forecasts. This allowed us to calculate expected values only for site data, rather than gridded data, greatly speeding up the analysis.}
For temperature, the expected value (but not the thresholded site forecast) was adjusted using the dry adiabatic lapse rate of -0.0098 \textdegree C/m to account for the difference between the grid point average altitude and the site altitude.

As described in Section \ref{blending}, we created two blended forecasts using either the NWP models only, or the full set of models.

\subsection{IMPROVER processing}\label{improver}

IMPROVER produces gridded forecasts over the Australian domain. Below we describe the main post-processing steps implemented by our IMPROVER operational configuration. More detail can be found in \cite{improver-rr} and \cite{fuzzy}.

\begin{enumerate}
	\item Linear interpolation from original lead times to hourly forecast.
	\item Regridding to 2400m Albers equal area grid using bilinear interpolation. (Note that this is in addition to the initial regridding of AIFS from the N320 grid as mentioned above.)
	\item Gridded lapse rate correction (temperature forecast only). Since the grid of the raw forecast is different to the IMPROVER output grid, we apply a correction to adjust for the difference between the average altitudes in each grid cell.
	\item Bias correction (applied to each ensemble member in the case of ensemble forecasts). A separate correction is applied for each grid point and lead time, based on the previous 30 days' history of forecast bias.  The historical forecasts are archived after the regridding step (or the lapse rate correction step, in the case of temperature). The ground truth for the correction is the gridded MSAS analysis. In the case of the ENS forecast, the bias is calculated from the ensemble mean forecast.
	\item Thresholding (ensemble forecast) or fuzzy thresholding (deterministic forecast) to produce a probabilistic forecast. IMPROVER represents probabilistic forecasts as a set of thresholds and their corresponding probabilities of exceedence.  For an ensemble forecast, thresholding simply means calculating the proportion of ensemble members that exceed the threshold. For a deterministic forecast, simple thresholding would yield a probability of either 0 or 1, which is not useful for reliability calibration. Instead, we use the fuzzy thresholding approach, where the predicted value is mapped to a number between 0 and 1 depending on how far it is from the threshold. Specifically, a parameter $r$ is chosen, representing the distance from the threshold within which we want to measure variation, and then for a deterministic forecast $f$ and a threshold $t$, the output $f^*$ of fuzzy thresholding is the number between 0 and 1 defined as follows
	\[
	f^* = \begin{cases}
	0, & x < t - r,\\
	(f - t + r) / (2r), & t - r \leq f \leq t + r,\\
	1, & x > t + r
	\end{cases}
	\]
	
	See \cite{fuzzy} for more details  on this technique.
	\item Neighbourhood processing with neighbourhoods of size $3 \times 3$ grid cells. This smooths the probabilistic forecast by convolving it with a constant-valued square filter. See \cite{roberts} for more detail.
	\item Recursive filtering. An additional spatial smoothing operation, similar to an exponential weighted average filter. See \cite{roberts} for more detail.
	\item Reliability calibration. Probabilities are adjusted using the reliability calibration method of \cite{flowerdew}. A piecewise-linear correction function is calculated as follows. For each threshold, the previous 30 days' history of forecasts are binned into 7 probability bins, and for each bin, the average forecast and observed probabilities of exceeding the threshold are calculated. The forecast and observed probabilities define the $x$ and $y$ coordinates, respectively, of the knot points of the piecewise linear correction function. The number of bins was chosen to give enough flexibility to model the calibration curve while avoiding overfitting (due to small sample size in some bins) and keeping the computation time feasible. The first bin contains only the value zero, and the remaining bins divide the interval $(0, 1]$ into equally-sized intervals. Our binning arrangement is similar to that used in \cite{rust}, who also use 7 bins in total, including two single-value bins at 0 and 1. The historical forecasts are archived after the thresholding step. Reliability calibration is calculated separately for each lead time, using the aggregate of all grid points. Thus (unlike the bias correction) the reliability calibration can address only global calibration problems, not those specific to particular locations or regions. Calculating separate reliability calibrations for each grid point would not be computationally feasible, and would also risk overfitting, given the short calibration period. As is the case for bias correction, the MSAS analysis is used as ground truth. 
\end{enumerate}

\subsection{Blending weights}\label{blending}
Forecast blending is the practice of combining outputs of multiple forecasts. In our case the blends are simple weighted averages of either the probability distribution outputs or the expected value outputs of post-processing. In order to investigate the benefits of adding AI models to a blend of NWP models, we compare two blended forecasts: the first uses only the two NWP models, ENS and HRES, and the second uses all three models. To supplement this analysis, we also consider the two other two-model blends, namely AIFS with HRES, and AIFS with ENS. In each case, the blending weights are chosen as follows. The data is split in half by valid time.\footnote{Valid time refers to the time the forecast is for.} Weights are fitted separately on each half of the data, and then applied to create the blended forecast for the other half of the data (this approach is often called 2-fold cross-validation). In this way, we avoid fitting the weights on the same data we use to measure model performance. In general, the performance of a weighting model is expected to be slightly better when applied to the same dataset used to optimise the weights, compared to an unseen dataset, because the weights can fit to random noise in the training dataset. Thus our cross-validation approach is intended to obtain a realistic indication of the model's performance in an operational setup, where the weights must be fitted on historical data, and then applied on previously-unseen data.

The ground truth for fitting the weights is the observation data at sites, and data is pooled so that weights are shared across all sites. To produce weights that vary smoothly over lead times, we use the following process. First, we calculate the optimal weights at each lead time to minimise the mean squared error. However, these weights vary greatly from hour to hour, partly due to diurnal patterns in the model biases, and partly due to random noise. To reduce overfitting and improve generalisation, we calculate a second set of smoothed weights. The smoothed weight for each model is a piecewise-linear function of lead time with 11 knots at lead times 0, 24, 48, ... 240. The function values at the knot points are chosen to minimise the MSE between the piecewise linear function and the individual hourly optimal weights, then normalised to sum to 1 over all models at each knot point (which guarantees that the interpolated values between knot points also sum to 1).

\begin{figure*}
\begin{center}
\includegraphics[width=15cm]{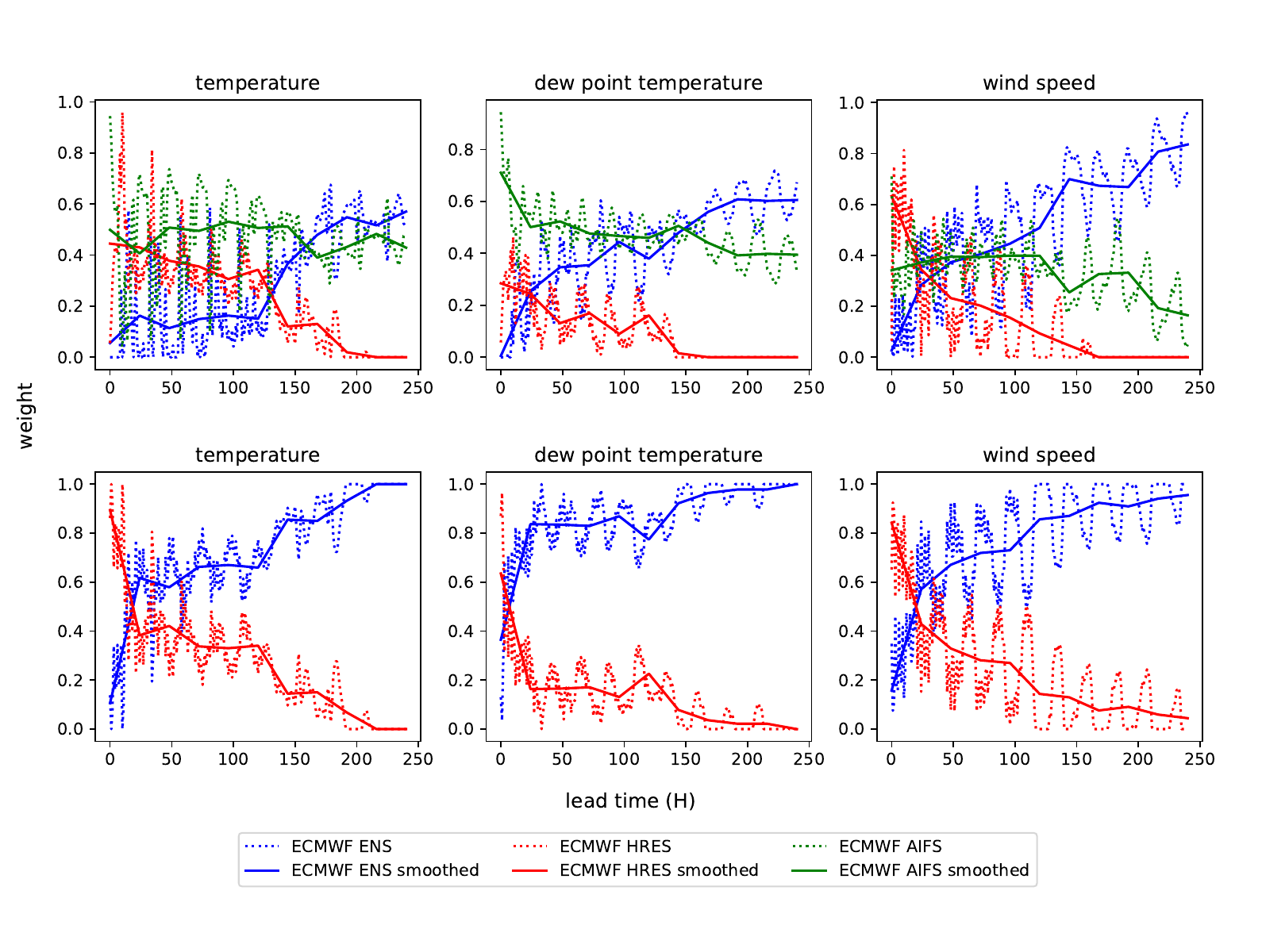}
\caption{Blending weights for the all-model blend (top row) and NWP-model blend (ENS and HRES only; bottom row). The dotted line shows the optimal weights per lead hour. The solid line shows a smoothed version where a piecewise-linear function is fitted to the optimal weights. These weights are calculated on the first half of the dataset, and are used to calculate the blended output for the second half.}
\label{fig:weights}
\end{center}
\end{figure*}

Figure \ref{fig:weights} shows the optimal weights when blending either all models, or only the two NWP models. (As described above, two sets of weights were fitted; the weights shown here are those fitted on the first half of the data.) For all variables, AIFS contributes significantly to the blend, and for temperature and dew point, it is generally the main component in the early part of the forecast period.

\section{Results}\label{results}

Our verification compares site-extracted forecast outputs with site observations, using the data described in Section \ref{data}. Note this is different to the ground truth used in the IMPROVER processing: since IMPROVER produces gridded forecasts, the bias correction and reliability calibration steps use the MSAS gridded analysis, as described in Section \ref{methods}\ref{improver}. 

As noted by \cite{aifs} and \cite{benbouallegue}, ML forecasts tend to have a smoother appearance than physics-based NWP, because they are based on optimising a loss function, which rewards predicting the expected value of the distribution, rather than simulating physical processes.  Additionally, graph neural networks, including the graph transformer architecture used in AIFS, are susceptible to a problem where, as the number of layers increases, message-passing between nodes increases similarity between the features of different nodes \citep{graphsmoothing}, which would contribute to visual smoothness of the gridded output. Figure \ref{fig:grid} compares gridded HRES and AIFS temperature forecasts before and after post-processing. The input AIFS is noticeably smoother and less detailed than HRES, although this is partly also due also to their different resolutions, \ang{0.25} vs \ang{0.1}.\footnote{For a comparison of AIFS vs the ECMWF integrated forecasting system (IFS) at the same resolution, see Figure 7 of \cite{aifs}. Similarly, Figure 3 in \cite{benbouallegue} compares the IFS with Pangu-Weather at equal resolution. In both examples, the AI forecasts are smoother than their NWP counterparts.} However, after post-processing, the AIFS forecast becomes more detailed and looks quite similar to HRES (although still a little smoother); in particular, post-processing is able to add realistic topographic features. Moreover, the fact that post-processed AIFS is slightly smoother than post-processed HRES does not seem to disadvantage it when assessing performance at sites: AIFS generally has better performance.

\begin{figure*}
\begin{center}
\includegraphics[width=15cm]{grid\_tempscreen}
\caption{Raw forecasts (top row) and post-processed expected value outputs from IMPROVER (bottom row) for temperature at lead time 18H and valid time 2024-06-15 06:00 UTC (16:00 Australian Eastern Standard Time). The left column is HRES and the right AIFS. Units are degrees Celsius. Note that the raw forecasts are on a latitude/longitude grid, while the post-processed forecasts use the Albers equal area projection. The diagonal artefacts near the edges in the post-processed forecasts occur because calibration is done against the MSAS analysis, which has a more limited spatial domain.}
\label{fig:grid}
\end{center}
\end{figure*}

Figure \ref{fig:msenwppost} shows forecast mean squared error by lead day before and after IMPROVER post-processing. For the raw ENS forecast, the quantity verified is the ensemble mean. Lead day 0 corresponds to lead hours 0-24 inclusive, lead day 1 to hours 25-48, et cetera.  Post-processing generally improves the raw forecasts, yielding gains of a few days of skill in some cases. For all variables, the three post-processed models achieve similar accuracy early in the forecast period. For temperature and dew point, after the first few lead days AIFS has a clear advantage over HRES, with the gap widening as lead time increases; both are outperformed by ENS. The strong performance of AIFS is likely because it is trained specifically to minimise the MSE objective, rather than aiming to model the most likely forecast trajectory like HRES. However, as with NWP models, the autoregressive nature of AIFS means that small errors are amplified over the forecast period, which causes AIFS (and HRES) to fall behind ENS in accuracy as the lead time increases.

It is well established that blending models generally improves the accuracy relative to the best individual forecast (see, for example, \cite{vannitsem}), and therefore it is a common approach in operational weather forecasts. Figure \ref{fig:mseblend} compares the two different blends, along with the post-processed models. To measure the difference between the two blends, we used the Diebold-Mariano test to calculate a confidence interval for the difference in MSE for each lead day. This test requires that the time series of differences 
be weakly stationary. Therefore, in order to avoid diurnal and spatial biases in the error, for each lead time we aggregated the data by valid day and calculated the MSE (including all sites and valid times for that valid day).\footnote{The aggregation is slightly different from the aggregation by lead day shown in Figure \ref{fig:msenwppost}, which simply combines all data for each lead day, without the intermediate step of taking the mean by valid day.} The test statistic was calculated on this daily series, using the implementation in the ``scores'' Python package \citep{scores} based on the method of \cite{hg}. Figure \ref{fig:blendcomparisonmse} shows the MSE difference with confidence interval. Overall, the blend including all three models is significantly better than the blend of only the two NWP models.

To more closely examine the ways the different models contribute to the blend, we also considered two additional blends: a blend of HRES and AIFS, and a blend of ENS and AIFS. Figures \ref{fig:mseallblends} and \ref{fig:crpsallblends} in the Appendix show the MSE and CRPS for the full array of blends (see also Tables \ref{tab:msetabletemp}, \ref{tab:msetabledewpoint}, \ref{tab:msetablewindspeed} and Tables \ref{tab:crpstabletemperature}, \ref{tab:crpstabledewpoint}, and \ref{tab:crpstablewindspeed} for the corresponding data). The blend of ENS and AIFS achieves very similar accuracy to the all-model blend. Interestingly, the blend of the only the two deterministic models, HRES and AIFS, also achieves similar accuracy to the all-model blend in the first few days of the forecast period. This offers the opportunity for operational centers to obtain high-quality short-term forecasts by using only deterministic forecasts, which are computionally much cheaper than ensembles. At later lead times the blends including ENS perform much more strongly, which is expected as the chaotic behaviour of the atmosphere makes deterministic forecasting difficult at these horizons.

\begin{figure*}
\begin{center}
\includegraphics[width=15cm]{mse\_nwp\_vs\_post}
\caption{Mean squared error by lead day for raw (dashed line) and post-processed (solid line) models ENS (blue), HRES (red) and AIFS (green), for temperature (left), dew point (middle) and wind speed (right).  The calculation includes only lead times that are present in both the raw and post-processed forecasts (that is, those that are multiples of 6 hours).}
\label{fig:msenwppost}
\end{center}
\end{figure*}

\begin{figure*}
\begin{center}
\includegraphics[width=15cm]{mse\_by\_day\_blends}
\caption{Mean squared error by lead day for post-processed models ENS (blue dashed), HRES (red dashed) and AIFS (green dashed) and blends (NWP models ENS and HRES, yellow; and all models, pink).  Note that the line for ENS is very close to that of the NWP-model blend, and partially hidden by it. The tabular data for these graphs can be found in Tables \ref{tab:msetabletemp}, \ref{tab:msetabledewpoint} and \ref{tab:msetablewindspeed} in the Appendix.}
\label{fig:mseblend}
\end{center}
\end{figure*}

\begin{figure*}
\begin{center}
\includegraphics[width=15cm]{blend\_comparison\_mse}
\caption{MSE difference between NWP-model blend and all-model blend. The difference is $\mathrm{MSE_{NWP}} - \mathrm{MSE_{all}}$, so positive values indicate that the all-model blend is better. The shaded region is the 95\% confidence interval.}
\label{fig:blendcomparisonmse}
\end{center}
\end{figure*}

\begin{figure*}
\begin{center}
\includegraphics[width=13cm]{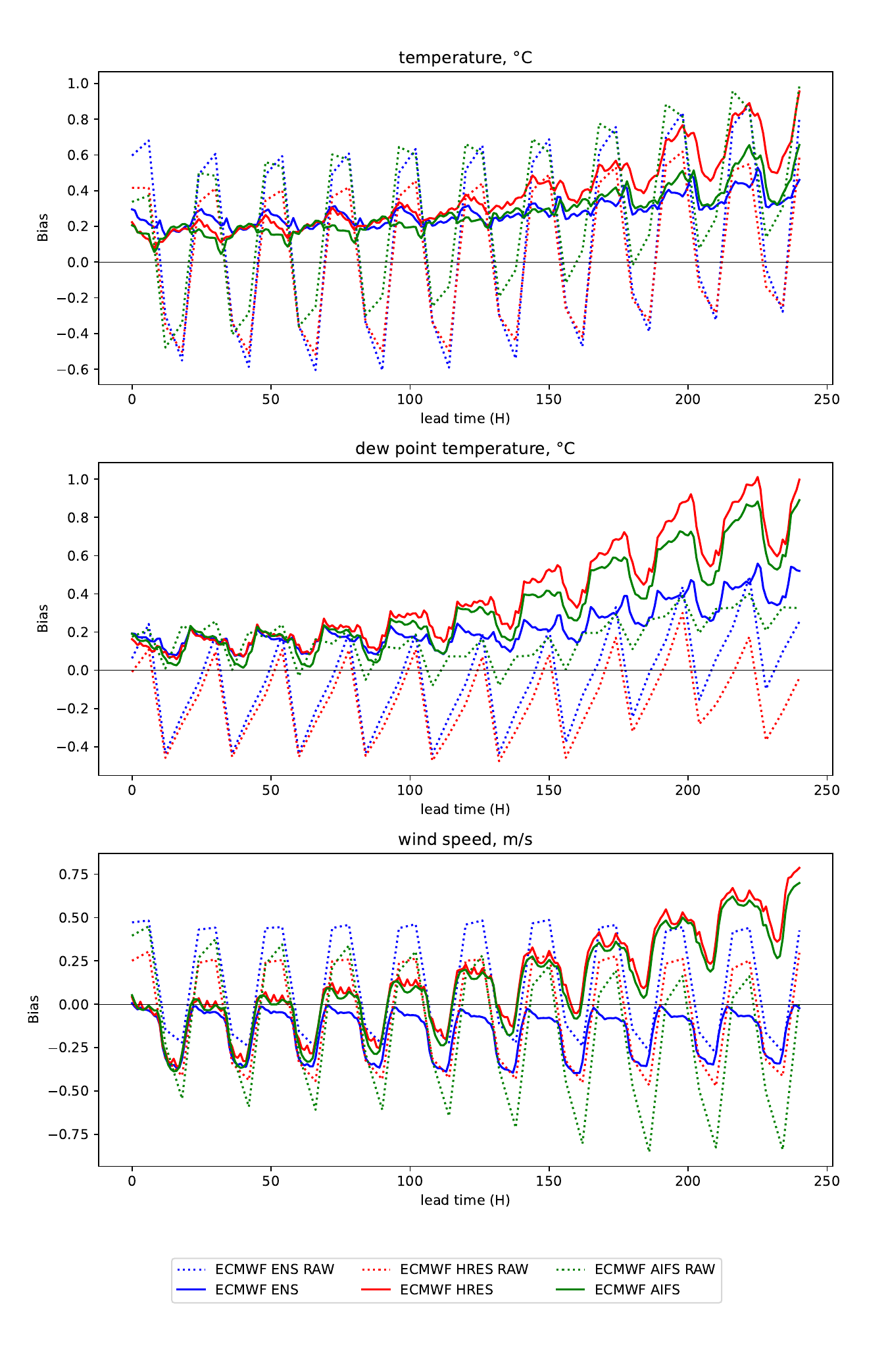}
\caption{Bias of raw (dotted line) and post-processed (solid line) models ENS (blue), HRES (red) and AIFS (green) over the forecast period, averaged over all observation sites. Note that the raw models have 6-hour lead time frequency, while the post-processed outputs have 1-hour frequency.}
\label{fig:bias}
\end{center}
\end{figure*}

Figure \ref{fig:bias} shows the bias (defined as forecast minus observation) of the various models at the observation sites. As with MSE, for the raw ENS forecast, bias is calculated for the ensemble mean. Bias correction significantly mitigates the biases present in the raw forecasts. However some diurnal patterns and constant biases remain.  This is to be expected because the IMPROVER bias correction is done relative to the MSAS analysis, which has its own biases relative to the site observation data used for verification (see Figure \ref{fig:msasbias} in the Appendix). We also observe trends in the bias over the forecast period.  This is somewhat surprising, given that the bias correction and reliability calibration are calculated separately for each lead time.  One possible explanation is that the calibration is based on a moving window of the past 30 days' history, so at longer lead times there is a greater lag between the validity time of the forecast and the most recent ground truth used in the calibration. This perhaps reduces the ability of the calibration to correct for both seasonal model biases (that is, those varying systematically by time of year), and for idiosyncratic errors in the currently-occuring weather system. An alternative approach could be to use a longer historical calibration dataset, containing a year or more of data, and calculate seasonally-varying biases. However, this requires a long history of model data, which is not available for new models like AIFS, and would also be more complex to implement operationally.

\begin{figure*}
\begin{center}
\includegraphics[width=15cm]{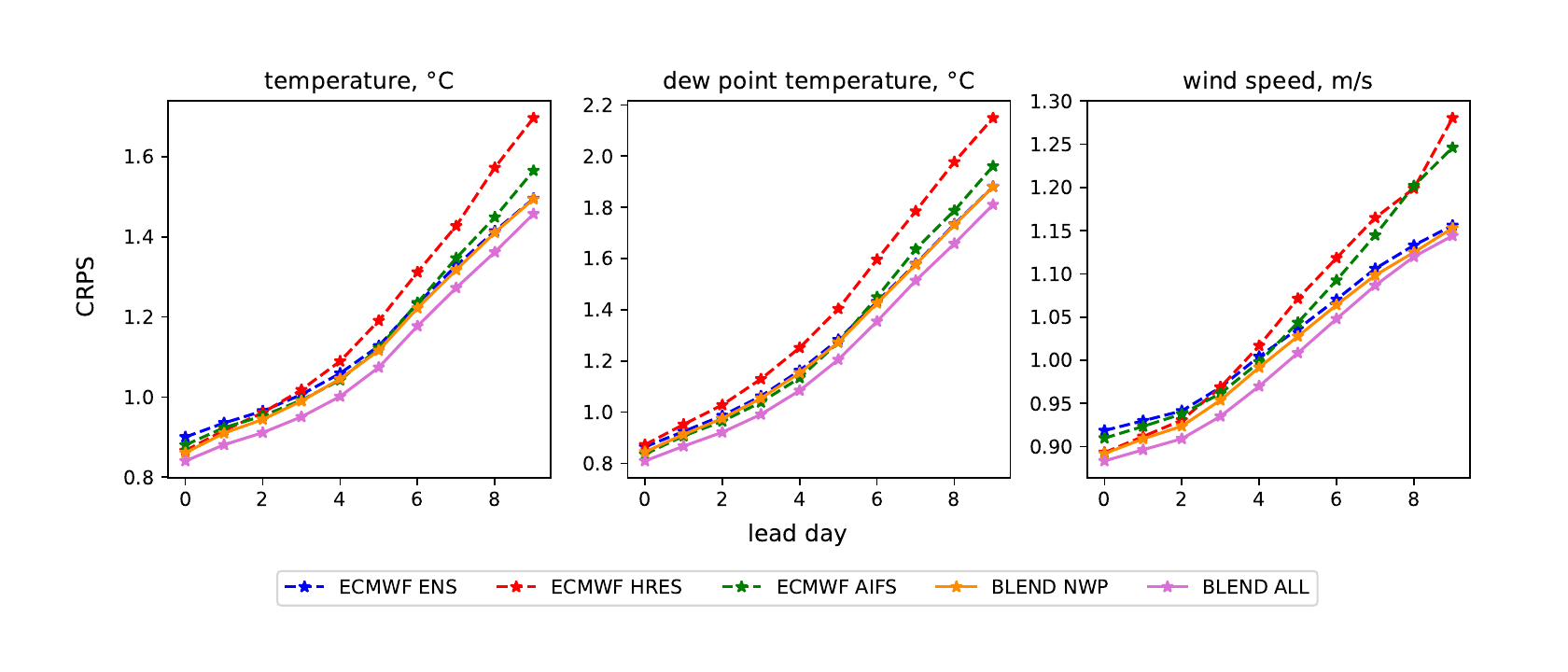}
\caption{Continuous ranked probability score (CRPS) by lead day for post-processed models ENS (blue dashed), HRES (red dashed), and AIFS (green dashed), and blends (NWP models ENS and HRES, yellow; and all models, pink) over the forecast period. The tabular data for these graphs can be found in Tables \ref{tab:crpstabletemperature}, \ref{tab:crpstabledewpoint} and \ref{tab:crpstablewindspeed} in the Appendix.}
\label{fig:crps}
\end{center}
\end{figure*}

IMPROVER produces probabilistic outputs individually for both deterministic and ensemble input forecasts. For evaluating their accuracy, we use the continuous ranked probability score (CRPS), which measures the error between the predicted probability distribution and the true outcome. The CRPS is defined as the integral of the squared distance between the forecast distribution and the true outcome. Formally, let $X$ be the random variable being forecast, $t :\to \mathrm{P}(X \leq t)$ the predicted cumulative distribution function for some data point, and $x$ the observed value. Then the CRPS for this observation is
\[
\int_{t=-\infty}^{t=\infty} (\mathrm{P}(X \leq t)  - \mathbbm{1}_{(t \geq x)})^2 dt
\]
where $\mathbbm{1}_{(t \geq x)}$ is the indicator function which is equal to 1 if $t \geq x$ and 0 otherwise. Smaller CRPS values indicate a better forecast. When evaluating the CRPS, we take the forecast CDF to be the piecewise-linear function obtained by linearly interpolating the probabilities of exceedence at the predicted thresholds. Figure \ref{fig:crps} shows the CRPS of the post-processed models and blends.  While in general the blend of the two NWP models performs very similarly to the most accurate individual NWP model, the blend of all three models yields a larger improvment. We used the Diebold-Mariano test to 
compare the two blends, following the same process as described above for the MSE comparison. The results are shown in Figure \ref{fig:blendcomparisoncrps}; as is the case with MSE, we see that the all-model blend is significantly better than the NWP-model blend.

\begin{figure*}
\begin{center}
\includegraphics[width=15cm]{blend\_comparison\_crps}
\caption{CRPS difference between NWP-model blend and all-model blend. The difference is $\mathrm{CRPS_{NWP}} - \mathrm{CRPS_{all}}$, so positive values indicate that the all-model blend is better. The shaded region is the 95\% confidence interval.}
\label{fig:blendcomparisoncrps}
\end{center}
\end{figure*}

Forecast reliability measures how well-calibrated a forecast is. Specifically, for a given threshold $t$, a forecast has good reliability if the conditional probability $p_o$ that an observation exceeds $t$, given that the forecast probability is $p_f$, is approximately equal to $p_f$. Figure \ref{fig:reliability} shows the reliability at the 12-hour lead time (UTC 0) at selected thresholds, roughly corresponding to the 10th, 50th and 90th percentiles of the overall distribution of observations (including all times of day). Note that only a few thresholds are shown here; IMPROVER calculates probabilities at a much larger set of thresholds as described in Section \ref{methods}. There is no clear trend in these results: all models and blends have similar performance. The deterministic models 
arguably appear a little better than the ensemble, and the blended models do not seem to offer much advantage. Reliability is still fairly good later in the forecast period: Figure \ref{fig:reliability228} in the Appendix shows the reliability at lead time 228 hours. It has been shown by \cite{ranjan} that blending probability forecasts does not preserve calibration; that is, even if each individual forecast is perfectly calibrated, in general the blend is not. Therefore, calibration could likely be further improved by recalibrating the blend. Functionality to implement the parametric recalibration proposed by Ranjan and Gneiting has recently been implemented in the IMPROVER software library, but is not currently part of the IMPROVER operational system used for this work. 

In Figure \ref{fig:distributions} we show the distributions of probabilistic forecasts at the 12-hour lead time, which allows 
us to evaluate the forecasts' sharpness. A sharp forecast is indicated by a deep U-shaped distribution graph, with many values close to 0 or 1, while a flatter graph indicates a less-confident forecast. The 3-model blend is in general less sharp than the 2-model blend, which is in turn less sharp than the sharpest NWP model, namely ENS. Again, this is expected for blended forecasts, and could be corrected with recalibration.

\begin{figure*}
\begin{center}
\includegraphics[width=15cm]{lt12\_reliability}
\caption{Reliability at the 12-hour lead time for post-processed models ENS (blue dashed), HRES (red dashed), and AIFS (green dashed), and blends (NWP models ENS and HRES, yellow; and all models, pink). Forecast probabilities are binned into 10 equal-width bins, and, for each bin, the average forecast and observed probability of exceeding the threshold is calculated. These values are linearly interpolated to produce the reliability curve. Bins having fewer than 10 data points are not plotted.}
\label{fig:reliability}
\end{center}
\end{figure*}

\begin{figure*}
\begin{center}
\includegraphics[width=15cm]{distribution\_lt12}
\caption{Distribution of binned probability forecasts at the 12-hour lead time for post-processed models ENS (blue dashed), HRES (red dashed), and AIFS (green dashed), and blends (NWP models ENS and HRES, yellow; and all models, pink). Forecasts have been binned into 10 equal-width bins. The $y$-axis uses the symmetric log 
scale, which is linear for values between 0 and 1, and logarithmic for larger values.}
\label{fig:distributions}
\end{center}
\end{figure*}

\section{Discussion}\label{discussion}

This study demonstrates that established statistical post-processing methods, originally designed for traditional Numerical Weather Prediction (NWP) models, are also effective when applied to forecasts from the Artificial Intelligence Forecasting System (AIFS). Using the Australian Bureau of Meteorology's operational IMPROVER system without modification, we observe comparable accuracy improvements in AIFS forecasts to those seen with NWP models. Notably, blending AIFS with traditional NWP models yields consistent gains in forecast skill for both expected value and probabilistic forecasts. These results indicate that AI-based forecasts can be integrated into operational systems using existing tools and workflows, thereby expanding the utility of these forecasts without the need for bespoke post-processing solutions.

A primary finding of this work is that statistical calibration via IMPROVER substantially improves the accuracy of AIFS forecasts. The gains are comparable to those achieved when calibrating traditional models, and in some cases, calibration adds upwards of a day of forecast skill. This reinforces our view that AI models, like their physics-based counterparts, benefit from correction of systematic errors using historical forecast performance. Importantly, the post-processing workflow required no changes, suggesting a high degree of generality in these methods.

The calibrated AIFS output also produces high-quality probabilistic forecasts. For temperature and dew point in particular, the post-processed AIFS shows CRPS values similar to those of the ensemble system (ENS) during the early part of the forecast period. This is notable given that AIFS is a deterministic model, and the result highlights the value of fuzzy thresholding and reliability calibration in extracting probabilistic information from deterministic model inputs.

Visual inspection of gridded output shows that post-processing adds realistic spatial structure to AIFS forecasts, compensating for the low resolution of the raw model. This is achieved by bias correcting against the MSAS analysis, which has much finer raw resolution than AIFS, and, in the case of temperature, by lapse-rate adjustment to downscale to the finer grid. Our results demonstrate that statistical techniques can improve not only  quantitative accuracy but also  the visual detail of the forecast fields -- an important element in operational weather forecasting.

The inclusion of AIFS in a blended forecast with NWP models consistently improves forecast accuracy across all variables (temperature, dew point, and wind speed). This result holds for both the deterministic mean squared error metric and probabilistic CRPS metric. The benefits of blending persist even when AIFS is individually less accurate than other models. This underscores the value of diversity in model formulation and suggests that AI models offer complementary strengths to traditional NWP models and enhance overall system performance when combined in this way.

These results provide a practical pathway for the adoption of AI-based weather forecasts by national meteorological centres. With AIFS now being officially supported by ECMWF, the demonstrated compatibility with existing post-processing systems means this forecast can be incorporated into current workflows without requiring customised infrastructure. Critically, blending allows for a flexible, low-risk approach to operational use: rather than relying entirely on an AI model, centres can integrate it alongside traditional NWP models, assigning weights according to performance and risk tolerance. This incremental adoption strategy enables institutions to harness the benefits of AI innovation while maintaining forecast robustness and continuity.

A limitation of the present work is that the verification period is quite short and does not represent all seasons. It is likely that the input models have seasonally-varying strengths and weaknesses, so it would be valuable to perform the analysis on a longer period. Future work could also consider extending these analyses to AIFS-CRPS \citep{aifscrps}, an ensemble version of AIFS trained to minimise CRPS. This may further improve the skill of blended forecasts. Additionally, recalibrating the probabilistic blends -- particularly in light of known issues with calibration preservation during blending -- may yield further improvements.

\clearpage
%%%%%%%%%%%%%%%%%%%%%%%%%%%%%%%%%%%%%%%%%%%%%%%%%%%%%%%%%%%%%%%%%%%%%
% ACKNOWLEDGMENTS
%%%%%%%%%%%%%%%%%%%%%%%%%%%%%%%%%%%%%%%%%%%%%%%%%%%%%%%%%%%%%%%%%%%%%
\acknowledgments

We wish to acknowledge ECMWF for making available data for the deterministic, ensemble, and AI forecasts; and the National Computing Infrastructure (NCI) Australia for providing the computing facilities used for this analysis. 
We are grateful to the UK Met Office for their contributions to the IMPROVER partnership.
We thank Timothy Hume for contributions to the IMPROVER implementation at the Bureau, and for feedback on this paper. We also thank Mengmeng Han and Nicholas Loveday for feedback on a draft of this work.

\datastatement

Data for the raw AIFS, ENS, and HRES forecasts is available from ECMWF via the MARS API. The open-source IMPROVER post-processing software is available from \url{https://github.com/metoppv/improver/}.

%  The data availability statement is where authors should describe how the data underlying 
%  the findings within the article can be accessed and reused. Authors should attempt to 
%  provide unrestricted access to all data and materials underlying reported findings. 
%  If data access is restricted, authors must mention this in the statement. See
%  {http://www.ametsoc.org/PubsDataPolicy} for more info.

%%%%%%%%%%%%%%%%%%%%%%%%%%%%%%%%%%%%%%%%%%%%%%%%%%%%%%%%%%%%%%%%%%%%%
% APPENDIXES
%%%%%%%%%%%%%%%%%%%%%%%%%%%%%%%%%%%%%%%%%%%%%%%%%%%%%%%%%%%%%%%%%%%%%
%
%% If only one appendix, use

\newpage
\appendix

\begin{figure*}
\begin{center}
\includegraphics[width=15cm]{mse\_by\_day\_blends\_full\_list}
\caption{Mean squared error by lead day for all post-processed NWP models and blends. Note that the BLEND ALL and BLEND ENS AIFS lines are almost identical.}
\label{fig:mseallblends}
\end{center}
\end{figure*}

\begin{table*}[h]
\caption{Temperature mean squared error by lead day, $(^{\circ}\mathrm{C})^2$, corresponding to Figure \ref{fig:mseblend} and Figure \ref{fig:mseallblends}.}\label{tab:msetabletemp}
\begin{center}
\begin{tabular}{lrrrrrrrrrr}
\toprule
lead day & 0 & 1 & 2 & 3 & 4 & 5 & 6 & 7 & 8 & 9 \\
forecast &  &  &  &  &  &  &  &  &  &  \\
\midrule
ECMWF ENS & 2.43 & 2.52 & 2.73 & 3.00 & 3.37 & 3.86 & 4.46 & 5.24 & 5.99 & 6.72 \\
ECMWF HRES & 2.43 & 2.61 & 2.87 & 3.23 & 3.73 & 4.48 & 5.38 & 6.41 & 7.80 & 9.11 \\
ECMWF AIFS & 2.23 & 2.40 & 2.58 & 2.82 & 3.16 & 3.75 & 4.68 & 5.61 & 6.51 & 7.65 \\
BLEND NWP & 2.39 & 2.51 & 2.71 & 2.99 & 3.36 & 3.85 & 4.45 & 5.23 & 6.00 & 6.74 \\
BLEND ALL & 2.17 & 2.30 & 2.47 & 2.69 & 3.02 & 3.53 & 4.20 & 4.96 & 5.71 & 6.57 \\
BLEND HRES AIFS & 2.17 & 2.31 & 2.47 & 2.70 & 3.03 & 3.59 & 4.38 & 5.24 & 6.25 & 7.49 \\
BLEND ENS AIFS & 2.19 & 2.32 & 2.48 & 2.70 & 3.02 & 3.53 & 4.21 & 4.96 & 5.70 & 6.56 \\
\bottomrule
\end{tabular}

\end{center}
\end{table*}

\begin{table*}[h]
\caption{Dew point mean squared error by lead day, $(^{\circ}\mathrm{C})^2$, corresponding to Figure \ref{fig:mseblend} and Figure \ref{fig:mseallblends}.}\label{tab:msetabledewpoint}
\begin{center}
\begin{tabular}{lrrrrrrrrrr}
\toprule
lead day & 0 & 1 & 2 & 3 & 4 & 5 & 6 & 7 & 8 & 9 \\
forecast &  &  &  &  &  &  &  &  &  &  \\
\midrule
ECMWF ENS & 2.73 & 3.11 & 3.54 & 4.09 & 4.88 & 5.91 & 7.30 & 8.89 & 10.62 & 12.44 \\
ECMWF HRES & 2.82 & 3.36 & 3.91 & 4.69 & 5.71 & 7.11 & 9.24 & 11.35 & 13.59 & 15.86 \\
ECMWF AIFS & 2.43 & 2.88 & 3.27 & 3.80 & 4.54 & 5.74 & 7.49 & 9.48 & 11.26 & 13.48 \\
BLEND NWP & 2.71 & 3.10 & 3.53 & 4.09 & 4.86 & 5.88 & 7.30 & 8.91 & 10.65 & 12.47 \\
BLEND ALL & 2.37 & 2.74 & 3.09 & 3.57 & 4.26 & 5.25 & 6.64 & 8.28 & 9.88 & 11.75 \\
BLEND HRES AIFS & 2.37 & 2.76 & 3.11 & 3.61 & 4.34 & 5.43 & 7.09 & 9.01 & 10.85 & 13.03 \\
BLEND ENS AIFS & 2.38 & 2.75 & 3.09 & 3.57 & 4.25 & 5.25 & 6.62 & 8.27 & 9.88 & 11.75 \\
\bottomrule
\end{tabular}

\end{center}
\end{table*}

\begin{table*}[h]
\caption{Wind speed mean squared error by lead day, $(\mathrm{m/s})^2$, corresponding to Figure \ref{fig:mseblend} and Figure \ref{fig:mseallblends}.}\label{tab:msetablewindspeed}
\begin{center}
\begin{tabular}{lrrrrrrrrrr}
\toprule
lead day & 0 & 1 & 2 & 3 & 4 & 5 & 6 & 7 & 8 & 9 \\
forecast &  &  &  &  &  &  &  &  &  &  \\
\midrule
ECMWF ENS & 2.77 & 2.83 & 2.94 & 3.17 & 3.46 & 3.70 & 3.96 & 4.25 & 4.45 & 4.64 \\
ECMWF HRES & 2.75 & 2.88 & 3.02 & 3.31 & 3.68 & 4.10 & 4.50 & 4.88 & 5.22 & 5.91 \\
ECMWF AIFS & 2.74 & 2.83 & 2.95 & 3.14 & 3.45 & 3.82 & 4.23 & 4.68 & 5.19 & 5.54 \\
BLEND NWP & 2.74 & 2.82 & 2.93 & 3.15 & 3.44 & 3.69 & 3.95 & 4.25 & 4.45 & 4.64 \\
BLEND ALL & 2.68 & 2.75 & 2.85 & 3.05 & 3.32 & 3.61 & 3.89 & 4.21 & 4.45 & 4.61 \\
BLEND HRES AIFS & 2.68 & 2.75 & 2.87 & 3.07 & 3.36 & 3.73 & 4.12 & 4.56 & 5.03 & 5.42 \\
BLEND ENS AIFS & 2.69 & 2.75 & 2.86 & 3.05 & 3.32 & 3.61 & 3.89 & 4.18 & 4.42 & 4.60 \\
\bottomrule
\end{tabular}

\end{center}
\end{table*}

\begin{figure*}
\begin{center}
\includegraphics[width=15cm]{msas\_bias}
\caption{Bias of MSAS analysis against site observations for each hour of the day in UTC time (positive values indicate that MSAS is higher). When comparing with Figure \ref{fig:bias}, note that the forecast basetime is at UTC 12, so lead time 
0 in Figure \ref{fig:bias} corresponds to hour 12 in this figure.}
\label{fig:msasbias}
\end{center}
\end{figure*}

\begin{figure*}
\begin{center}
\includegraphics[width=15cm]{crps\_by\_day\_blends\_full\_list}
\caption{Continuous ranked probability score by lead day for all post-processed NWP models and blends. Note that the BLEND ALL and BLEND ENS AIFS lines are almost identical. }
\label{fig:crpsallblends}
\end{center}
\end{figure*}

\begin{table*}[h]
\caption{Temperature continuous ranked probability score by lead day, $^{\circ}\mathrm{C}$, corresponding to Figure \ref{fig:crps} and Figure \ref{fig:crpsallblends}.}\label{tab:crpstabletemperature}
\begin{center}
\begin{tabular}{lrrrrrrrrrr}
\toprule
lead day & 0 & 1 & 2 & 3 & 4 & 5 & 6 & 7 & 8 & 9 \\
forecast &  &  &  &  &  &  &  &  &  &  \\
\midrule
ECMWF ENS & 0.90 & 0.94 & 0.96 & 1.01 & 1.06 & 1.13 & 1.23 & 1.33 & 1.41 & 1.50 \\
ECMWF HRES & 0.87 & 0.92 & 0.96 & 1.02 & 1.09 & 1.19 & 1.31 & 1.43 & 1.57 & 1.70 \\
ECMWF AIFS & 0.88 & 0.92 & 0.95 & 0.99 & 1.04 & 1.12 & 1.24 & 1.35 & 1.45 & 1.57 \\
BLEND NWP & 0.86 & 0.91 & 0.94 & 0.99 & 1.05 & 1.12 & 1.22 & 1.32 & 1.41 & 1.49 \\
BLEND ALL & 0.84 & 0.88 & 0.91 & 0.95 & 1.00 & 1.07 & 1.18 & 1.27 & 1.36 & 1.46 \\
BLEND HRES AIFS & 0.84 & 0.88 & 0.92 & 0.96 & 1.01 & 1.09 & 1.20 & 1.31 & 1.42 & 1.55 \\
BLEND ENS AIFS & 0.85 & 0.89 & 0.92 & 0.95 & 1.00 & 1.07 & 1.18 & 1.27 & 1.36 & 1.46 \\
\bottomrule
\end{tabular}

\end{center}
\end{table*}

\begin{table*}[h]
\caption{Dew point continuous ranked probability score by lead day, $^{\circ}\mathrm{C}$, corresponding to Figure \ref{fig:crps} and Figure \ref{fig:crpsallblends}.}\label{tab:crpstabledewpoint}
\begin{center}
\begin{tabular}{lrrrrrrrrrr}
\toprule
lead day & 0 & 1 & 2 & 3 & 4 & 5 & 6 & 7 & 8 & 9 \\
forecast &  &  &  &  &  &  &  &  &  &  \\
\midrule
ECMWF ENS & 0.86 & 0.92 & 0.99 & 1.06 & 1.16 & 1.28 & 1.43 & 1.58 & 1.73 & 1.88 \\
ECMWF HRES & 0.87 & 0.95 & 1.03 & 1.13 & 1.25 & 1.40 & 1.60 & 1.78 & 1.98 & 2.15 \\
ECMWF AIFS & 0.84 & 0.91 & 0.96 & 1.04 & 1.13 & 1.27 & 1.45 & 1.64 & 1.79 & 1.96 \\
BLEND NWP & 0.84 & 0.91 & 0.98 & 1.05 & 1.15 & 1.27 & 1.43 & 1.58 & 1.73 & 1.88 \\
BLEND ALL & 0.81 & 0.87 & 0.92 & 0.99 & 1.08 & 1.21 & 1.35 & 1.51 & 1.66 & 1.81 \\
BLEND HRES AIFS & 0.82 & 0.88 & 0.94 & 1.01 & 1.11 & 1.24 & 1.41 & 1.60 & 1.76 & 1.93 \\
BLEND ENS AIFS & 0.80 & 0.86 & 0.92 & 0.99 & 1.08 & 1.20 & 1.35 & 1.51 & 1.66 & 1.81 \\
\bottomrule
\end{tabular}

\end{center}
\end{table*}

\begin{table*}[h]
\caption{Wind speed continuous ranked probability score by lead day, $\mathrm{m/s}$, corresponding to Figure \ref{fig:crps} and Figure \ref{fig:crpsallblends}.}\label{tab:crpstablewindspeed}
\begin{center}
\begin{tabular}{lrrrrrrrrrr}
\toprule
lead day & 0 & 1 & 2 & 3 & 4 & 5 & 6 & 7 & 8 & 9 \\
forecast &  &  &  &  &  &  &  &  &  &  \\
\midrule
ECMWF ENS & 0.92 & 0.93 & 0.94 & 0.97 & 1.00 & 1.04 & 1.07 & 1.11 & 1.13 & 1.16 \\
ECMWF HRES & 0.89 & 0.91 & 0.93 & 0.97 & 1.02 & 1.07 & 1.12 & 1.16 & 1.20 & 1.28 \\
ECMWF AIFS & 0.91 & 0.92 & 0.94 & 0.96 & 1.00 & 1.04 & 1.09 & 1.14 & 1.20 & 1.25 \\
BLEND NWP & 0.89 & 0.91 & 0.92 & 0.95 & 0.99 & 1.03 & 1.06 & 1.10 & 1.12 & 1.15 \\
BLEND ALL & 0.88 & 0.90 & 0.91 & 0.94 & 0.97 & 1.01 & 1.05 & 1.09 & 1.12 & 1.14 \\
BLEND HRES AIFS & 0.88 & 0.90 & 0.91 & 0.94 & 0.98 & 1.02 & 1.07 & 1.12 & 1.17 & 1.22 \\
BLEND ENS AIFS & 0.89 & 0.90 & 0.92 & 0.94 & 0.97 & 1.01 & 1.05 & 1.09 & 1.12 & 1.14 \\
\bottomrule
\end{tabular}

\end{center}
\end{table*}

\begin{figure*}
\begin{center}
\includegraphics[width=15cm]{lt228\_reliability}
\caption{Reliability at the 228-hour lead time for post-processed models ENS (blue dashed), HRES (red dashed), and AIFS (green dashed), and blends (NWP models ENS and HRES, yellow; and all models, pink). Forecast probabilities are binned into 10 equal-width bins, and, for each bin, the average forecast and observed probability of exceeding the threshold is calculated. These values are linearly interpolated to produce the reliability curve. Bins having fewer than 10 data points are not plotted.}
\label{fig:reliability228}
\end{center}
\end{figure*}

%% If more than one appendix, use \appendix[<letter>], e.g.,

%\appendix[A] 

%% Appendix title is necessary! For appendix title:

%\appendixtitle{Title of Appendix}

%%% Appendix section numbering (note, skip \section and begin with \subsection)
%
% \subsection{First primary heading}

% \subsubsection{First secondary heading}

% \paragraph{First tertiary heading}

\clearpage
\clearpage
\newpage
%%%%%%%%%%%%%%%%%%%%%%%%%%%%%%%%%%%%%%%%%%%%%%%%%%%%%%%%%%%%%%%%%%%%%
% REFERENCES
%%%%%%%%%%%%%%%%%%%%%%%%%%%%%%%%%%%%%%%%%%%%%%%%%%%%%%%%%%%%%%%%%%%%%
% Make your BibTeX bibliography by using these commands:
\bibliographystyle{ametsocV6}
\bibliography{references}

\end{document}